\documentclass[11pt]{article}
\usepackage{hyperref}
\pdfoutput=1

\begin{document}
\title{New (Internal) Wave Generation - \\ Laboratory Experiments}
\author{Manikandan Mathur$^{[1]}$, Matthieu Mercier$^{[2]}$, \\
Thierry Dauxois$^{[2]}$, Thomas Peacock$^{[1]}$ \\ \\
$[1]$ Dept. of Mech. Engg., MIT, USA. \\
$[2]$ Laboratoire de Physique, ENS Lyon, France.}

\maketitle

\begin{abstract}
In this fluid dynamics video, we demonstrate the experimental generation of 
various internal wave fields using a novel wave generator. Specifically, uni-directional 
internal wave beams and vertical modes $1$ and $2$ are generated and visualized using 
Particle Image Velocimetry. Further details and analysis of these experiments 
can be found in \cite{newwave2009}. 
\end{abstract}

The first one minute of this three-minute video shows the working and assembly of the wave generator. 
In all the experiments shown, quantitative velocity field measurements were performed using Particle 
Image Velocimetry.

The 1:00-1:40 section of the video presents the generated wave field for wave beam experiments. The phase of the
oscillating plates in these experiments travels upwards (0:54-1:03), and this ensures that the emitted wave field propagates 
predominantly downwards. The first experiment (1:04-1:24), performed in a linear density stratification with 
buoyancy frequency $N=0.85$ rad/s, corresponds to $\omega=0.22$ rad/s, and produces a wave beam that propagates
at an angle $\theta = \sin^{-1}\frac{\omega}{N}=15^{\circ}$ with respect to the horizontal. In the second experiment 
(1:25-1:40), the forcing frequency is increased to $\omega=0.6$ rad/s (with the value of $N$ the same as before), 
and we observe a wave beam that propagates at $\theta=\sin^{-1}\frac{\omega}{N}=45^{\circ}$. The generated and the 
reflected (off the bottom of the tank) wave beams interfere to form a striking array of vortices.

The 1:40-2:15 section of the video shows how one can excite vertical mode 1 internal waves using the wave generator. Forcing the horizontal
velocity in the shape of mode-1, we observe a traveling internal wave that spans the entire height of the fluid. The final section 
demonstrates the generation of mode-2 internal waves in a similar manner. These experiments, which correspond to $N=0.85$ rad/s and 
$\omega=0.6$ rad/s, prove that the generator can excite distinct modes with remarkable efficiency.  

High and low resolution versions of the video can be found at
\href{http://ecommons.library.cornell.edu/handle/1813/14111}{final\_high\_res.mpeg} (1.38 GB) and
\href{http://ecommons.library.cornell.edu/handle/1813/14111}{final\_low\_res.mpeg4} (9.47 MB), respectively.

\end{document}